\documentclass[fleqn,10pt]{wlscirep}
\usepackage[utf8]{inputenc}
\usepackage[T1]{fontenc}

\usepackage{multirow}

\newcommand{\etal}{\textit{et al.}}
\newcommand{\RM}[1]{\MakeUppercase{\romannumeral #1{}}}

\title{Deep Learning-Based Quantification of Pulmonary Hemosiderophages in Cytology Slides}

\author[1,2,*]{Christian Marzahl}
\author[1]{Marc Aubreville}
\author[3]{Christof A. Bertram}
\author[4]{Jason Stayt}
\author[5]{Anne-Katherine Jasensky}
\author[3]{Florian Bartenschlager}
\author[3]{Marco Fragoso-Garcia}
\author[6]{Ann K. Barton}
\author[7]{Svenja Elsemann}
\author[8]{Samir Jabari}
\author[2]{Jens Krauth}
\author[1]{Prathmesh Madhu}
\author[2]{Jörn Voigt}
\author[4]{Jenny Hill}
\author[3]{Robert Klopfleisch}
\author[1]{Andreas Maier}

\affil[1]{Pattern Recognition Lab, Friedrich-Alexander-Universität Erlangen-Nürnberg, Erlangen, Germany}
\affil[2]{Research and Development, EUROIMMUN Medizinische Labordiagnostika AG, Lübeck, Germany}
\affil[3]{Institute of Veterinary Pathology, Freie Universität Berlin, Germany}
\affil[4]{VetPath Laboratory Services, Ascot,Western Australia}
\affil[5]{Laboklin GmbH und Co. KG, Bad Kissingen, Germany}
\affil[6]{Equine Clinic, Freie Universität Berlin, Berlin, Germany}
\affil[7]{Department of Neurosurgery, Universitätsklinikum Erlangen, Erlangen, Germany}
\affil[8]{Institute of Neuropathology, Friedrich Alexander University Erlangen-Nürnberg, Erlangen, Germany}

\affil[*]{c.marzahl@euroimmun.de}

\begin{abstract}
\emph{Purpose}: Exercise-induced pulmonary hemorrhage (EIPH) is a common syndrome in sport horses with negative impact on performance. Cytology of bronchoalveolar lavage fluid by use of a scoring system is considered the most sensitive diagnostic method. Macrophages are classified depending on the degree of cytoplasmic hemosiderin content. The current gold standard is manual grading, which is however monotonous and time-consuming. 
\emph{Methods}: We evaluated state-of-the-art deep learning-based methods for single cell macrophage classification and compared them against the performance of nine cytology experts and evaluated inter- and intra-observer variability. Additionally, we evaluated object detection methods on a novel data set of 17 completely annotated cytology whole slide images (WSI) containing 78,047 hemosiderophages. 
\emph{Results}: Our deep learning-based approach reached a concordance of 0.85, partially exceeding human expert concordance (0.68 to 0.86, $\mu$=0.73, $\sigma$ =0.04). Intra-observer variability was high (0.68 to 0.88) and inter-observer concordance was moderate (Fleiss’ kappa = 0.67). Our object detection approach has a mean average precision of 0.66 over the five classes from the whole slide gigapixel image and a computation time of below two minutes. 
\emph{Conclusion}: To mitigate the high inter- and intra-rater variability, we propose our automated object detection pipeline, enabling accurate, reproducible and quick EIPH scoring in WSI.

\end{abstract}
\begin{document}

\flushbottom
\maketitle
%
%
\thispagestyle{empty}

\section*{Introduction}

Patients with pulmonary hemorrhage (P-Hem) suffer from repeated bleeding into the lungs, which can result in dyspnea and if untreated, may have life threatening consequences~\cite{ahmad2019morbidity}. There are various causes which lead to P-Hem, including drug abuse, premature birth, leukaemia, autoimmune disorders and immunodeficiencies~\cite{maldonado2009haemosiderin, van1992pulmonary, golde1975occult, martinez2017diffuse, kahn1987diagnosis}. In this paper, we focus on a special subtype of P-Hem called exercise-induced pulmonary hemorrhage (EIPH) in animals. Specifically, EIPH affects racing horses and causes reduced athletic performance~\cite{morley2015exercise,hinchcliff2005association,birks2003exercise, hinchcliff2015exercise}. 
The gold standard for diagnosis of P-hem in humans and equine animals is to perform cytology of bronchoalveolar lavage fluid 
(BALF)~\cite{hoffman2008bronchoalveolar} using a scoring system as explained by Golde \etal~\cite{golde1975occult}.
The red blood cells of the bleeding are degraded into an iron-storage complex called hemosiderin by alveolar macrophages. Hemosiderin-laden macrophages are called hemosiderophages.
Prior to microscopic evaluation, the cells extracted by the BALF are stained with Perls’
Prussian Blue~\cite{depecker2015comparison} or Turnbull’s Blue~\cite{denk1989romeis} in order to visualise the iron pigments contained in the hemosiderin. According to the commonly used scoring system (Macrophages hemosiderin score) by Golde \etal~\cite{golde1975occult}, alveolar macrophages can be distinguished into five grades depending on their hemosiderin content. This scoring system is based on the fact that the hemosiderin concentration correlates directly with the degree of prior P-hem.

The macrophages hemosiderin score is determined on cytological specimens, which can be digitalised using a whole slide scanner resulting in a whole slide image (WSI). One of the main issues with manual counting of hemosiderophages in digital microscopy - just like in traditional light microscopy - is that it is a laborsome and time-consuming task. Secondly, and more importantly these images are commonly subject to inter- and intra-observer variability. Additionally, there is the problem that hemosiderin absorption is a continuous process which is mapped to a discrete grading system. 
To our knowledge, no previous research has investigated the use of end-to-end, deep learning-based object detection methods for the multi-class problem of pulmonary hemorrhage on WSI.
In particular, no study to date has examined the inter- and intra-observer variability for hemosiderophage classification, which is crucial when comparing human performance to algorithmic approaches. Especially, since there is no measurable ground truth available and therefore the consistency of the ground truth annotation by an expert is unknown. 
In this work, the main objective is to develop an overarching deep learning based system for the analysis of whole slide EIPH images. This includes the detection and classification of hemosiderophages in an accurate, efficient, explanatory and reliable manner. 

The major contributions of this paper are as follows: Firstly, we created the largest published data set of fully annotated EIPH images, containing 78,047 single cell annotations by a pathology expert. Secondly, we conducted an analysis of the inter- and intra-observer variability for the classification of single hemosiderophages (CoSH) by multiple experts and deep learning based methods. Thirdly, we developed a custom network architecture dedicated for the purpose of multi-class whole-slide analysis (MCWSA).

This results in a deployable object detection system for WSI EIPH images which can process gigapixel images in under two minutes on a  modern GPU and is freely available for research purposes. 

\section*{Related Work}

To date, the topic of hemosiderophage classification and quantification has not been solved using computer vision methods. However, there have been numerous studies in the past decades with the goal of detecting cells, nuclei and mitosis figures for multiple modalities like digital  fluorescence microscopy and histopathology~\cite{waithe2019object,baykal2019modern,aubreville2018field}. Historically, these methods started as hand-crafted low-level feature extraction methods~\cite{lowe1999object,ojala1996comparative,dalal2005histograms}. With the recent advent of deep learning based techniques~\cite{maier2019gentle} these methods transitioned into modern end-to-end optimised object detection algorithms like Faster-RCNN~\cite{ren2015faster}, SSD~\cite{liu2016ssd} or RetinaNet~\cite{Lin2017ICCV}. Their underlying end-to-end optimisation approach is the foundation of their success in object detection challenges for natural images like PASCAL VOC~\cite{everingham2010pascal} and MS COCO~\cite{lin2014microsoft} where no classical approach could outperform a modern deep learning based object detection method~\cite{zou2019object} since 2014.   
The aim of object detection algorithms is to predict the bounding box as well as a class for multiple objects irrespective of the scale or partial occlusion of the objects. These methods have generated state of the art results in the fields of pedestrian-, face- and car-detection and are used in state of the art autonomous vehicles as well as the interpretation of satellite images~\cite{lin2014microsoft,everingham2010pascal,mundhenk2016large}. Regarding the field of medical object detection, the review by Litjens \etal~\cite{litjens2017survey}  reveals that no one had implemented deep learning based object detection methods for the evaluation of medical images as of 2017. In contrast, they mention that sliding window approaches in combination with a deep learning based classification network or U-Net-like segmentation architectures~\cite{ronneberger2015u} are being commonly used. The frequent use of U-Net in particular is quite remarkable since segmentation provides no means of separating touching or overlapping objects and these methods highly rely on post-processing steps for the task of separation. Additionally, in the case of U-Net, the architectures are more computationally complex due to their encoder-decoder architecture. Moreover, these networks require a pixel-wise annotation mask for obtaining better results, which is time-consuming compared to the relatively simple and fast creation of bounding box annotations needed for object detection methods. Ferlaino  \etal~\cite{ferlaino2018towards} have used deep learning based object detection on fully annotated multiclass WSI. They use RetinaNet~\cite{Lin2017ICCV} for nuclei detection and a separate network for nuclei classification which is not end-to-end trainable.

Modern object-detection algorithms can be categorised into two major categories: a) single stage algorithms and b) two stage algorithms. In single-stage algorithms, the task of detection and classification is solved in one single step, examples are YOLO~\cite{redmon2016you}, SSD~\cite{liu2016ssd} or RetinaNet~\cite{Lin2017ICCV}. While in two stage algorithms, the task of detection is solved by the use of a region proposal network (RPN)~\cite{ren2015faster} in the first stage and then classified using a additional network in a subsequent stage. While two stage detection algorithms are more accurate in general, the single stage methods yield the better ratio of accuracy and inference speed~\cite{huang2017speed}. This trade-off between speed and accuracy is crucial when analysing WSI with billions of pixels.
In this paper, we will use RetinaNet as a starting point for analysing EIPH on WSI because its architecture is straightforward, easy to modify and adapt for WSI analysis.

\section*{Material}

Our research group built a data set of 17 cytological slides of equine bronchoalveolar lavage fluid. Slides had been prepared by cytocentrifugation and stained for iron content with Prussian Blue (n=10) or Turnbull's Blue (n=7) which result in identical colour pattern. Digitalisation of the glass slide was performed using a linear scanner (Aperio ScanScope CS2, Leica Biosystems, Germany) at a magnification of 400$\times$ (resolution: $ 0.25 ~\frac{\mu m}{px})$. Finally, the slides were completely annotated and scored by a veterinary pathologist. All bronchoalveolar lavage fluids were obtained from horses with clinical signs of lower respiratory tract disease with written informed consent forms from the owners and taken from routine diagnostic services. Therefore, no animal was harmed for the construction of this data set. Additionally, individual case histories were not considered in the present study. Using the open source software solution SlideRunner~\cite{aubreville2018sliderunner}, we were able to build a database that includes the annotations for each hemosiderophage on the slides with their corresponding grade. This was done by first annotating all pulmonary macrophages and afterwards classifying them into their corresponding grade. The scoring system for hemosiderophages was introduced by Golde \etal~\cite{golde1975occult} and consists of five classes: It starts with zero (no intracytoplasmic blue coloured pigment) and go up to four (cell filled with haemosiderin; dark blue throughout cytoplasm). The final score is calculated by the method of M. Y. Doucet and L. Viel~\cite{doucet2002alveolar} which is an adaptation of Golde ~\etal~\cite{golde1975occult} to be used for horses. In this scoring system, three hundred alveolar macrophages were first graded from zero to four, then the total number per grade was divided by three and multiplied with the corresponding grade. Afterwards, the resulting Total Hemosiderin Score (THS) ranges from zero to four hundred. If the score is higher than seventy-five then the diagnosis pulmonary hemorrhage is considered to be confirmed. The completely annotated data set consists of seventeen slides and covers an area of 1,266 mm$^{2}$($\mu$ = 74mm$^{2}$, $\sigma$  = 9mm$^{2}$) containing  78,047 labeled cells ($\mu$ = 4,591, $\sigma$  = 3,389)(see Table~\ref{tabDataSetOverview}) which makes it the largest published data set of hemosiderophages and one of the largest of WSI. This novel data set allows us to perform object detection on whole hemosiderophages slides, for the first time. 
To take into account that not all slides contain  hemosiderophages of grade three and four, we used the same 14 slides to train and validate. However, we used the upper half of each image for training and the lower half for validation in order to prevent over-fitting. Three separate slides were selected as hold out test set slides. 

\subsection*{Single cell inter- and intra-observer variability} 

To evaluate the human inter- and intra-observer variability for single cell classification, we extracted two test sets containing 1000 cells each. For test set \RM{1}, the images were randomly selected among the labelled cells resulting in a representative distribution. Test set \RM{2} contains 1000 cells with a balanced distribution of 200 cells per grade.

\begin{table}
\def\arraystretch{1.2}
\setlength{\tabcolsep}{0.5em}
\centering
\caption{Data set statistics for each fully annotated WSI. Plus their total number of alveolar macrophages / hemosiderophages, the number of cells for each grade and theri corresponding mean grade and standard deviation. Horizontal twin-line separates the train/validation set from the test set. }
\label{tabDataSetOverview}
\begin{tabular}{|l|r|r|r|r|r|r|r|r|r|r|c|}
\hline
\multirow{2}{2pt}{File} & \multicolumn{1}{c|}{Staining} & \multicolumn{1}{c|}{Total} & \multicolumn{1}{c|}{Score} & \multicolumn{7}{c|}{ Count of Cells by Grade} \\
&  & Cells &  & 0 & 1 & 2 & 3 & 4 & $\mu$ & $\sigma$ \\
\hline

01\_EIPH & Prussian & 4446  & 126 & 1013 & 1782 & 1218 & 348 & 85 & 1.26 & 0.96 \\ 
02\_EIPH & Prussian & 12812 & 72  & 5084 & 6203 & 1450 & 64 & 11 & 0.72 & 0.68 \\ 
03\_EIPH & Prussian & 6325  & 37  &  4295 & 1697 & 330 & 3 & 0 & 0.37 & 0.58 \\ 
04\_EIPH & Prussian & 5448  & 63  &  2551 & 2379 & 508 & 10 & 0 & 0.63 & 0.66 \\ 
05\_EIPH & Prussian & 2489  & 34  &  1754 & 634 & 99 & 2 & 0 & 0.34 & 0.55\\ 
06\_EIPH & Turnbull & 2992  & 41  &  1908 & 933 & 148 & 3 & 0 & 0.41 & 0.59   \\ 
07\_EIPH & Turnbull & 1073  & 235 &  48 & 127 & 352 & 495 & 51 & 2.35 & 0.91   \\ 
08\_EIPH & Turnbull & 924   & 67  &  471 & 290 & 160 & 3 & 0 & 0.67 & 0.76 \\ 
09\_EIPH & Turnbull & 4752  & 216 &  568 & 1053 & 932 & 1446 & 753 & 2.16 & 1.27   \\ 
10\_EIPH & Prussian & 10385 & 208 &  592 & 2131 & 4037 & 3098 & 527 & 2.08 & 0.96   \\ 
12\_EIPH & Prussian & 5751  & 59  &  2839 & 2452 & 435 & 25 & 0 & 0.59 & 0.65    \\ 
13\_EIPH & Turnbull & 1112  & 35  &  767 & 302 & 43 & 0 & 0 & 0.35 & 0.55   \\ 
14\_EIPH & Turnbull & 968   & 43  &  637 & 252 & 70 & 8 & 1 & 0.43 & 0.67  \\ 
15\_EIPH & Prussian & 3143  & 39  &  1995 & 1062 & 81 & 5 & 0 & 0.39 & 0.55   \\ 
\hline
\hline
11\_EIPH & Prussian & 1841 & 148 &  283 & 553 & 859 & 131 & 15 & 1.48 & 0.86   \\ 
16\_EIPH & Prussian & 6491 & 87  &  2611 & 2509 & 984 & 363 & 24 & 0.87 & 0.89   \\ 
17\_EIPH & Turnbull & 7095 & 133 &  1639 & 2566 & 1818 & 1066 & 6 & 1.33 & 0.99    \\ 

\hline
\end{tabular}
\end{table}

\section*{Methods}

The research was carried out in accordance with the Code of Ethics of the World Medical Association (Declaration of Helsinki) and the guidelines of the institutions conducting the experiments.

The aim of this work was to develop and compare algorithmic approaches for predicting the hemosiderophage score of WSI. 
In order to assess how challenging the classification of single hemosiderophages (CoSH) is, we investigated methods considering the single cell labels as a classification problem. In our first approach a distinct whole-numbered class was assigned to each cell. During the second approach, however, decimal numbers were assigned to each cell. We then compared the results with human performance. Additionally, we present methods for multi-class WSI analysis (MCWSA). Here, we adopted state of the art deep learning based object detection and regression approaches. We used a support vector machine-based regression method to draw a baseline. To compensate the sparse cell distribution, we introduced a novel quad-tree based sampling approach to train the object detection networks. 

\subsection*{Human Performance Evaluation}

In order to compare our algorithmic approaches with human recognition performance, we investigated accuracy and reproducibility of nine cytology experts. We divided them into three groups according to their qualification and experience with BAL cytology. Each group contains three participants: 
\begin{itemize}
\item (E)xpert: Veterinary pathologists or clinician with high degree of experience in BAL cytology.  
\item (P)rofessional: Professional clinician or pathologist with basic experience in BAL cytology. 
\item (B)eginner: General skills in cytology, but no experience with BAL cytology in particular. 
\end{itemize}

Each of the nine cytology experts was asked to classify two thousand cells from the single cell test set \RM{1} and \RM{2}. We did not set a time limit to perform this task. In order to measure the intra-observer variability, they were asked to classify all cells again two weeks after the first round. The participants were instructed to perform classification according to the methods published by Doucet \etal\cite{doucet2002alveolar}

\subsection*{Sampling Strategy} \label{SamplingStrategy}

For deep neural networks, it is beneficial to be trained with equally distributed labelled examples. As shown in Table~\ref{tabDataSetOverview}, cell grade 3 and 4 rarely occur on some of the WSI. For example, file 14 includes only one grade 4 and eight grade 3 hemosiderophages. This means that with an image size of 35,999$\times$34,118 pixels and random sampling with a patch size of 1024$\times$1024 pixels, the chance to sample the grade 4 cell is just 0.08\% percent. 
       
\subsubsection*{Two stage cluster sampling strategies}\label{SamplingStrategyTwoStage}

For this sampling strategy, we clustered all cells from one WSI on the basis of their grade. For training, we randomly selected one of those clusters and chose one of the cells within that cluster by chance. Then a patch is randomly shifted in the direct proximity of that cell and the area is sampled for training.        

\subsubsection*{Generic Quad tree sampling strategies}

\begin{figure}[hbt!]
\includegraphics[width=1\textwidth]{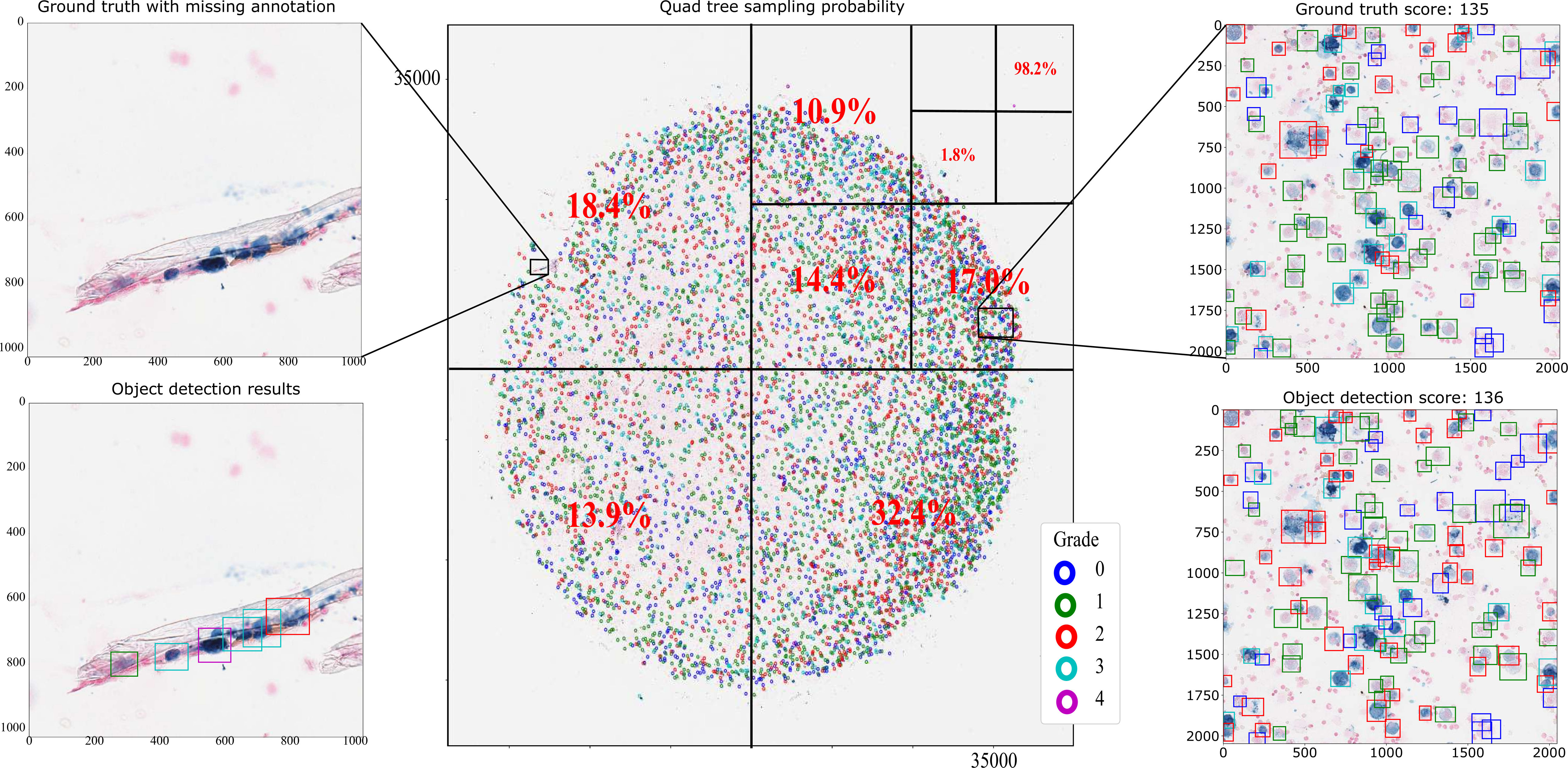}
\caption{ Left: This patch with  hemosiderophages was overlooked by the human expert which created the ground truth but correctly annotated by RetinaNet.  
Centre: Example for the sampling strategy on image 17\_EIPH Turnbull blue with 7095 cells. We can see a high sampling probability for the node with the only grade four cell. All cells are marked as dots. 
Right: Object detection result for an region of the image 17\_EIPH Turnbull blue with there ground truth top and the predictions bottom}
\label{figDetectionResults}
\end{figure}

We developed a novel sampling strategy for microscopy images based on a quad-tree in order to consider the probability of occurrence of cells as well as their neighbouring cells (see Fig.~\ref{figDetectionResults} center). 
At each level of the quad-tree (depth of the tree can be customised), we saved the cells, their corresponding sampling probability and their grade. As seen in Fig.~\ref{figDetectionResults} (center), at each level of the quad-tree, we have up to four nodes. One constraint for the tree while it is being created is that there must be at least three hundred cells in each node. One other option would be that the size of the final node must be identical to the training patch size (e.g. 1024$\times$1024 pixels). In contrast to the sampling strategy used in two stage cluster sampling strategies, we can sample at nodes without any cells by defining a minimum probability. To train our networks, we created a quad-tree with a depth of three. To create a training sample, we randomly traversed the quad-tree according to the sampling probability of the cells. Fig.~\ref{figDetectionResults} visualises this novel sampling approach. At the first level, the image is divided into four nodes with the sampling probabilities of 35.3\%, 32.4\% , 13.9\% and 18.4\% (clockwise). In this example, the top right node was selected by chance and was traversed further. This process was repeated until the final node at level three was reached and one patch was extracted for training.    

\subsection*{Single Cell Classification (CoCH)}

The hemosiderophages score is based on a subjective calculation method in which each cell in one region of the WSI is assigned one of five grades (ranging from zero to four). However, this grading system does not reflect the biological nature since there is a continuous gain of iron in the hemophages as opposed to a stepwise rise (as the quantized score indicates). To take this continuous increase into account, we propose a regression based cell score estimation. We then compared the result to the classification approach which mimics the human scoring system.  

\subsubsection*{Classification} \label{SingleCellClassification}

For the cell based classification task, we used a compact ResNet-18 Architecture~\cite{He:2016ib} pre-trained on ImageNet~\cite{russakovsky2015imagenet} with a fully connected two layer classification head and a final softmax activation. The cells used for training and validation were extracted according to the algorithm described in the previous generic quad tree based sampling strategy paragraph.
The Network was trained in two stages with the Adam optimiser and a maximal learning rate schedule of 0.01. Categorical cross entropy was used as the loss function. First, we trained only the classification head for three epochs, afterwards we fine-tuned the complete network for an additional twenty epochs until convergence was reached.

\subsubsection*{Regression}

\begin{figure}[hbt!]
\includegraphics[width=1\textwidth]{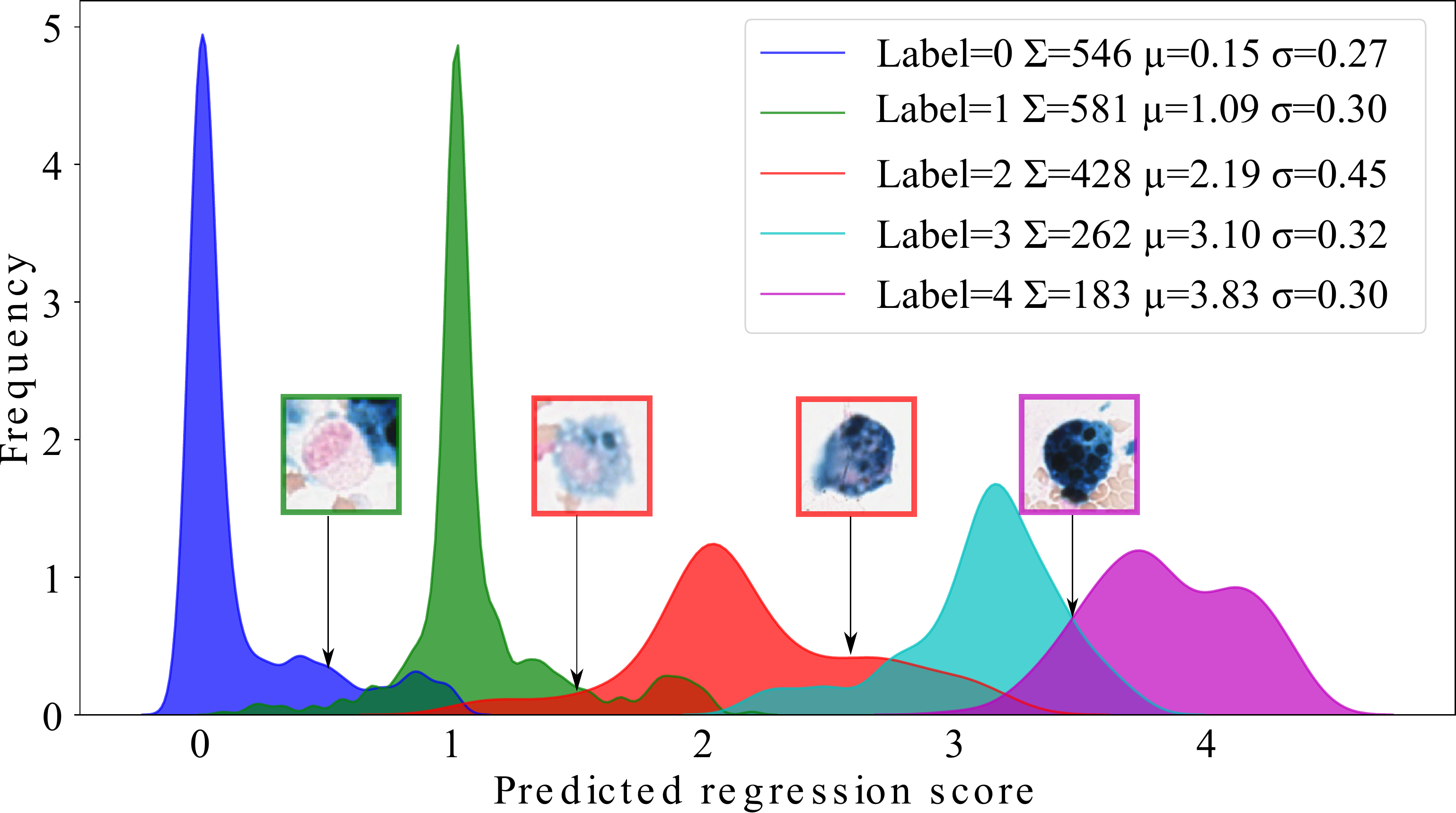}
\caption{Cell based regression results on the test data set visualised as a density histogram for the predicted scores. As an example, both cells in the middle are labelled with grade two and the regression model assigned very different scores to both, which is also clearly sensible from the visual appearance of the cell. }
\label{figDensityRegression}
\end{figure}

As stated, the hemosiderin absorption is a continuous process which is mapped to a discrete grading system. To take his continuity into account, we developed a network with a regression head and a final scaled sigmoid activation which predicts continues values in range of -0.5 to 4.5 to compensate the implementation instability for sigmoid activation's close to zero and one. The main focus of the experiment was to estimate the intra-grade confusion and increase the human interpretability of the results.   
This modification enables the network to predict decimal values between any two grades given that the cell has features of those two grades, which is not possible with a classification approach (see Fig. ~\ref{figDensityRegression}).
The network and training schedule is as described in the single cell classification paragraph. The mean squared error was used as loss function. 

\subsection*{Object detection based WSI score estimation (MCWSA)} \label{MCWSA}

\begin{figure}[hbt!]
\centering
\includegraphics{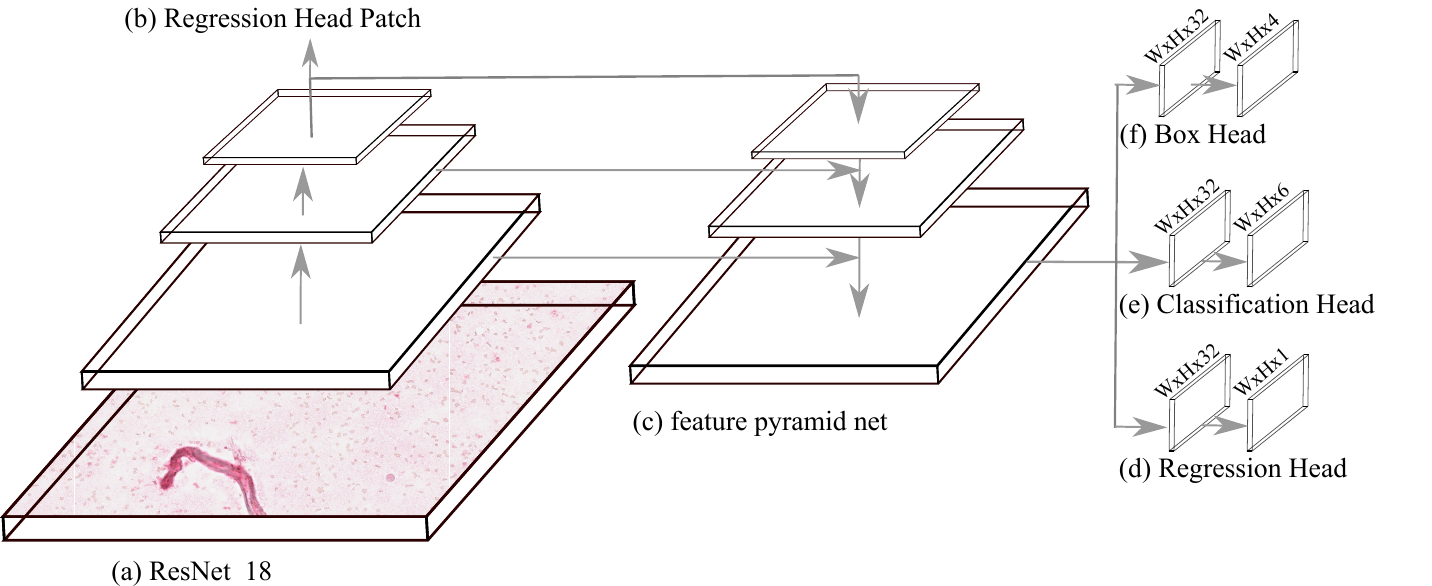}
\caption{Object detection and score prediction based on RetinaNet. a) ResNet-18 is used as input network for the c) Feature Pyramid Network~\cite{lin2017feature} to generate rich,  multi-scale features. The features ResNet-18 extracted from the patch are used for a direct regression based score estimation. d) Predicts a regression based score for each cell, e) classifies the cell into the five grades and background. f) Is used for regressing from anchors boxes to ground-truth bounding boxes. }
\label{figArchitecture}
\end{figure}

Besides investigating pure classification performance on single cells where the coordinates are previously known, the actual task in diagnostics is the estimation of scores on complete WSI or subparts thereof. Object detection networks mimic human expert behaviour by first detecting and classifying the cells and calculating the score afterwards. 
One object detection approach with a good accuracy-speed trade-off is RetinaNet~\cite{Lin2017ICCV} a single, unified network composed of a backbone network for feature extraction (see Fig. \ref{figArchitecture} a). A feature pyramid network (FPN)~\cite{lin2017feature} is built on top of the feature extractor to generate rich, multi-scale features by combining low-resolution with semantically strong features and high-resolution with semantically weak features (see Fig. \ref{figArchitecture} c). On each layer of the FPN, a classification- and a regression subnet are called to make predictions (see Fig. \ref{figArchitecture} d,e). The classification head predicts the probability of the target object's presence at each spatial position for each anchor. Anchors are defined by the scale and aspect ratio to match the targeted objects on each spatial position. To compensate for the class imbalance, focal loss~\cite{Lin2017ICCV} is used for training. The bounding box regression subnet (see Fig. \ref{figArchitecture} f) is generally built in a similar fashion as the classification head but is trained with smooth L1 loss and will predict four coordinates (x-offset, y-offset, width, height) for each box if a corresponding anchor box exists.     




We have modified the RetinaNet architecture in three significant ways to further optimise it for hemosiderophage WSI analysis.
First, we added an additional regression head which predicts the hemosiderophages score for each hemosiderophage (see Fig. \ref{figArchitecture} f). This was intended to increase the human interpretability of the results. As loss function for the cell based regression head, mean squared error is used. Second, to utilize the features extracted from the RetinaNet backbone, we fit an additional regression head on top of the ResNet-18 feature extractor for patch wise hemosiderophages score prediction. This process is further described in the later section  deep learning based regression and visualised in Figure \ref{figArchitecture} (b). Mean squared error is used as loss function for the patch based regression head. The total loss for training our network is calculated by Formula (Eq. \ref{equationTotalLoss}) where $c$ specifies the ground-truth grade, $\gamma$ is the tuneable focusing parameter, $\alpha _{t}$ the class imbalance weighting factor, $p_{t}$ is the model’s estimated probability for the class with grade c = 1 and x,y the arbitrary shapes. The network was trained with the Adam optimiser by using a maximal learning rate of 0.001 for 100 epochs until convergence was reached.
Additionally, to minimise the number of anchors and therefore further optimise the architecture towards inference speed we only used the 32$\times$32 feature map from the FPN. This was motivated by the fact that anchors of higher feature map sizes did not fit the small cell sizes and are limited in their total number.




\begin{equation}
\begin{split}
\operatorname{TotalLoss}(x, y, p_{t}, c) = &  \quad -\alpha _{t}(1 - p_{t})^{\gamma}log(p_{t})) 
\\ & + \frac{1}{n} \sum_{i}^{n} \left\{\begin{array}{ll}{0.5\left(x_{i}-y_{i}\right)^{2},} & {\text { if }\left|x_{i}-y_{i}\right|<1} \\ {\left|x_{i}-y_{i}\right|-0.5,} & {\text { otherwise }}\end{array}\right. 
\\ & + \frac{1}{n} \sum_{i=1}^{n}\left(c_{i}-\hat{c}_{i}\right)^{2}
\\ & + \frac{1}{n} \sum_{i=1}^{n}\left(c_{i}-\hat{c}_{i}\right)^{2}
\end{split}
\label{equationTotalLoss}
\end{equation}

For comparison, we additionally tested Faster-RCNN~\cite{ren2015faster} with a ResNet-50 backbone and SSD~\cite{liu2016ssd} with MobileNetV2 as provided by Huang \etal~\cite{huang2017speed}. Both networks were trained with the Adam optimiser and a learning rate of 0.0001 for 100 epochs until convergence was reached. All networks were trained with random rotation, horizontal and vertical flips, but no intensity augmentations. This was appropriate since a shift in intensity could alter the cell grade. 

\subsection*{Estimation based on image patch regression}

Direct estimation of the hemosiderophages score by using an image patch based regression approach is an alternative method if the bounding box illustration is not required. Furthermore, an image patch based regression approach could be used to find regions of interest efficiently even with standard computer vision methods which we will discuss in the following two methods for a regression based score estimation. While the first one uses a support vector machine (SVM), the second is an adaptation of the RetinaNet architecture. The goal of the regression based algorithm is to predict the grading score in range the from zero to four on a image patch and average the results for a total WSI.        

\subsubsection*{Support vector machine}

In order to set a computational inexpensive baseline for the task of estimating a hemosiderophages score, we trained a support vector machine with an Radial Basis Function (RBF) kernel and a convexity value of 0.1. These parameters were found by a grid search for the kernel and complexity parameter. As features we used the extracted histograms of a hundred patches per WSI with the sampling strategy described before . 

\subsubsection*{Deep learning based regression} \label{DeepLearningBasedRegression}

To estimate the hemosiderophages score with a deep learning based method we used the features extracted from RetinaNet and added two fully connected layers and a sigmoid activation for the regression head (see Fig. \ref{figArchitecture} b). To compensate the numerical instability for sigmoid activation's close to zero and one and in order to enable a sigmoid activation to predict scores up to grade four we scaled the sigmoid activation to a range from -0.5 and 4.5. The deep learning based regression network was trained as a part of our RetinaNet based object detection pipeline described in section object detection based WSI score estimation.

\section*{Results}

All experiments were run on a Linux workstation with a NVIDIA Quadro P5000 graphics card. The average calculation time for the object detection task was 101 seconds per whole slide image. The code for all experiments is available online and implemented in pytorch~\cite{paszke2017automatic} with fast.ai. The trained model can be downloaded freely and utilised with the open source software SlideRunner~\cite{aubreville2018sliderunner}.

\subsubsection*{Object Detection Evaluation}

\textit{Average Precision} (AP) was originally introduced in the 2007 PASCAL VOC challenge~\cite{everingham2010pascal} and is commonly used to assess object detection performance. AP is the average detection precision under different recalls and mean Average Precision (mAP) is the average over all five grades. 



\subsection*{Cell Classification (CoCH)} \label{CoCHResult}

\begin{figure}[hbt!]
\includegraphics[width=1\textwidth]{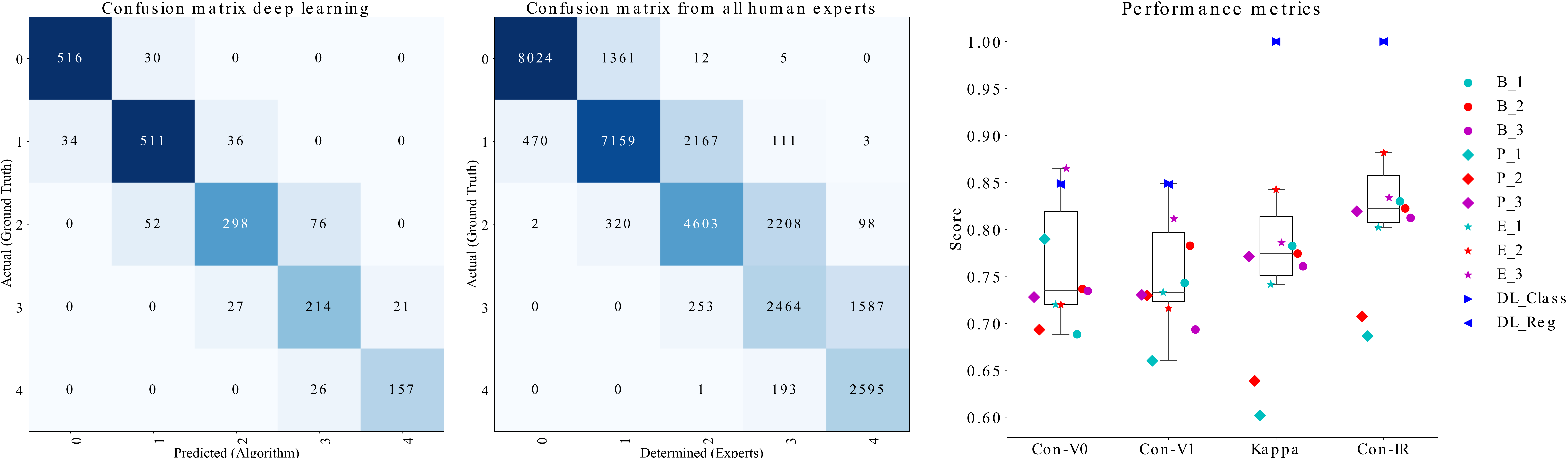}
\caption{From left to right: Confusion matrix for the automatic single cell classification results; Accumulated confusion matrix for all human experts; On the right the performance metrics diagram visualise results for the concordance with the ground truth for trail one and two (Con-V0, Con-V1). Additionally, the intra-rater concordance (Con-IR) and Cohens Kappa is shown. [B=Beginner, P=Professional, E=Expert, DL=Deep Learning approach.]}
\label{PerformanceMetrics}
\end{figure}

As stated above, we conducted an assessment of expert classification performance for comparison and to set a baseline. Comparing human experts and the deep learning classification pipeline, we found only offsets by one class by the deep learning system, whereas human expert disagreement was generally higher, especially for the higher grades 2, 3 and 4.In this categories disagreement in grade was significant for some cases (see Fig.~\ref{PerformanceMetrics}). Concordance with the ground truth data was 85\% for both automatic methods, whereas the human experts scored in range of 69-86\% ($\mu$=74, $\sigma$=5) for the first session (V0) and 66-81\% ($\mu$=73, $\sigma$=4) for the second session (V1). Illustrating that we were able to reach human expert-level concordance with the cell based regression and classification approach. The intra-observer variability ranges from 68-88\%($\mu$=79, $\sigma$=6) with a mean Cohen's kappa score of 0.74. The inter-observer Fleiss' kappa score is 0.67 at the first session (V0) and 0.68 at the second (V1). For the first session (V0) the f1 score per grade is 0 = 0.86 ($\sigma$=0.08), 1 = 0.74 ($\sigma$=0.08), 2 = 0.62 ($\sigma$=0.11), 3 = 0.50 ($\sigma$=0.16) and 4 = 0.68 ($\sigma$=0.21) and the second session (V1) 0 = 0.87 ($\sigma$=0.07), 1 = 0.73 ($\sigma$=0.07), 2 = 0.60 ($\sigma$=0.09), 3 = 0.47 ($\sigma$=0.14) and 4 = 0.61 ($\sigma$=0.28). 
The process of classifying two thousand cells took each expert roughly two hours while the deep learning approach took five seconds.
The human expert classification accuracy leads to a hypothetically mAP in the range of 0.57 (concordance 0.68) to 0.74 (concordance 0.86) with a mean of 0.60 (concordance 0.73) under the precondition that all cells are detected exactly as in the ground truth.
The ground truth mean score for the 2000 cells is 147 which was predicted by both deep learning approaches with a margin of 1 whereas the human's experts have a mean score error off -15 with a standard deviation of 12. The results are visualised in left sub figure of Fig.~\ref{figHeatMap}.

\subsection*{Object Detection (MCWSA)}

\begin{figure}[hbt!]
\includegraphics[width=1\textwidth]{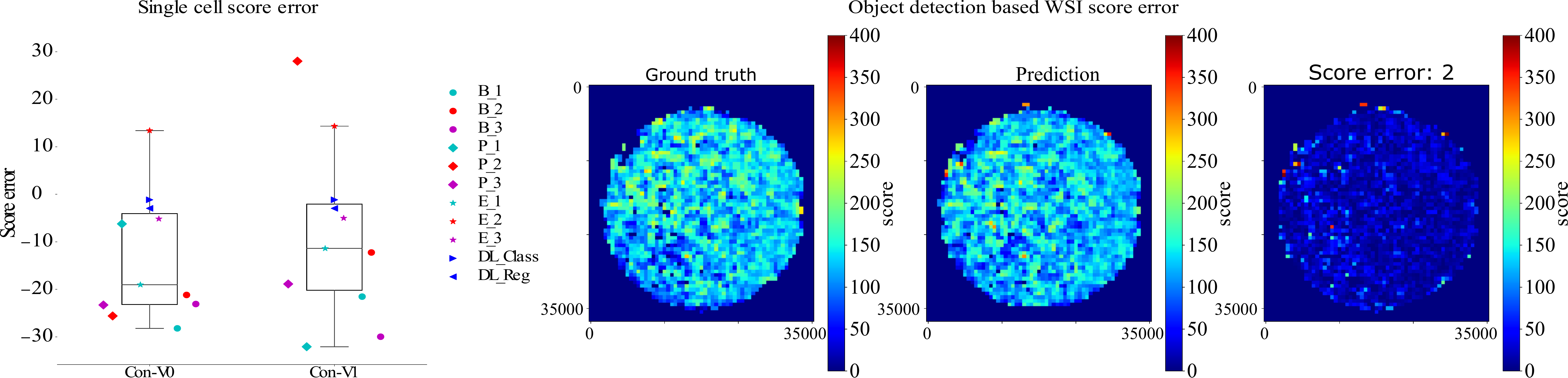}
\caption{The left diagram visualise the regression error for the single cell classification task. The three remaining figures show the object detection results from test set (slide \#17) on 1049 patches of size 1024$\times$1024. Ground truth (left), predictions (middle) and error (right). Large errors appearing at outer circle boundary can be explained by missed cell annotations.}
\label{figHeatMap}
\end{figure}

Our object detection approach showed a mean average precision (mAP) of 0.66 ($\sigma$=0.18, IoU=0.5) over the three test set WSI with a total of 3518 patches and 15,427 cells. Table~\ref{tabImplementationComparison} shows the results per WSI over all tested networks with a maximal mAP of 0.66 reached by multiple approaches.
The average error for cell-based grade score is 9 ($\sigma$  = 24) and is calculated by taking the absolute difference of all ground truth cell grades and the predicted grades. For better understanding a patch-wise analysed  WSI is shown in Fig.~\ref{figHeatMap}. 

The comparison of the three sampling strategies revealed a good overall convergence for the two-stage cluster sampling strategy (mAP 0.66)  and the quad tree sampling strategy (mAP 0.66), while completely random sampling showed very slow convergence to a lower mAP of 0.28.

\begin{table}
\def\arraystretch{1.2}
\setlength{\tabcolsep}{0.5em}
\centering
\caption{Comparison of multiple object detection architectures with their corresponding backbone, number of parameters, accuracy, score error and average inference speed per test WSI. We incrementally increased the number of channels and convolutional layers in our implementation until the accuracy converged against 0.66. Additionally the deep learning based regression and support vector machine error is shown for comparison.}
\label{tabImplementationComparison}
\begin{tabular}{|c|c|c|c|c|c|c|c|c|}
\hline
Architecture  & Backbone  & Parameter & mAP\_50 & Score Error & Inference speed \\
\hline
Ours  & RN-18   & 11.434.555            & 0.64   &  15  & 101s  \\ 
Ours  & RN-18   & 11.987.739            & 0.65   &  13  & 101s  \\ 
Ours  & RN-18   & 13.683.675            & 0.66   &  9  & 103s  \\ 
Ours  & RN-18   & 22.625.439            & 0.66   &  9  & 111s  \\ 
RetinaNet & RN-18 & 19.729.755          & 0.66   &  9  & 111s  \\ 
RetinaNet & RN-34  & 29.837.915         & 0.66   &  9  & 142s \\ 
RetinaNet & RN-50   & 36.288.347        & 0.66   &  8  & 258s  \\ 
SSD & MobileNetV2   &  13.871.354       & 0.61   &  21  & 105s  \\ 
Faster-RCNN & RN-50   & 128.383.642     & 0.66   &  7  & 305s  \\ 
\hline
\hline
SVM &  RBF-Kernel & / & / & 21 & 65s \\  
DL-Regression &  RN-18 & 11.704.897 & / & 19 & 92s \\  

\hline
\end{tabular}
\end{table}

\subsection*{Patch Regression}

As stated before, we evaluated two approaches to predict the grade score directly without additionally predicting bounding boxes and compared the results with our object detection based approach and the ground truth. The bounding box based approach produced the best results with an error of 9 compared to the deep learning based regression approach with 19 and the classical support vector based method with 21 as shown in (see Table~\ref{tabImplementationComparison}) bottom.    

\section*{Discussion and Outlook}

We demonstrated that the task of classifying hemosiderophages into the corresponding grading system as proposed by Golde \etal~\cite{golde1975occult} is not only monotonous and time-consuming but also highly subjective. This is highlighted by the observed high inter- and intra-observer variability and a moderate inter-rater reliability of agreement which strongly suggests that a discrete grading system has its limitations for the quantification of pulmonary hemosiderophages. This is an interesting topic for future work. Additionally, human experts who showed a tendency to assign grades below the reference grade were occasionally off by two grades. On the other hand, there was no obvious difference between the performance of the three defined groups of participants with different degree of experience with BAL cytology. In this paper, we proposed a single cell-based classification and regression system (CoCH) with a performance comparable to human experts in order to overcome this grading limitation. Plus, in contrast to the human experts, the classification and regression approaches showed both explanatory and reproducible outcomes while having an extremely high processing speed. However, the CoCH algorithm has the limitation that hemosiderophage cells had to be annotated by a human expert for further classification. Unfortunatelly, there is currently no true gold standard methods such as chemical measurement of iron content which ,of course, would be highly beneficial to validate our deep learning methods~\cite{hoffman2008bronchoalveolar}.

Scince manual scoring of P-Hem has some limitations, we proposed the use of computerised quantification. This could lead to a scoring with promising results regarding accuracy, reproducibility and inference speed. We have presented that even with a perfect detection rate at a human level classification, the mAP is less than 0.74. Based on this data set, this defines an upper baseline for human and algorithmic approaches, which was almost reached by the streamlined object detection pipeline based on the RetinaNet-Architecture (MCWSA). Patch-based regression approaches did not achieve the accuracy of object-based methods as a consequence of their susceptibility to blue coloured artefacts. 
The introduction of the quad-tree based sampling strategy leads to more stable and better results in the beginning of the training process but ends up with results similar to the two stage cluster based sampling method. Furthermore, besides the investigated intra-observer variablity for single cell classification, the regions of the WSI that had been completely missed to be annotated by the human expert (see Fig. \ref{figDetectionResults} top left) were often identified correctly by our method (see Fig. \ref{figDetectionResults} bottom left). To reduce the possibility that regions of the WSI are missed by the annotating human expert, an interactive augmented annotation method that was trained on already annotated WSI could be introduced. This interactive annotation process could further increase the quality of annotated WSI by highlighting areas of the WSI where human annotations and the deep learning based predictions strongly diverge (Fig.~\ref{figHeatMap}). Furthermore, this interactive annotation method could be used to decrease the amount of required human interactions for annotating WSI by first creating a preliminary result which is then checked by the experts. This process should be closely monitored in order to refrain from introducing a bias towards accepting the deep learning based predictions plus further research is required regarding reliability. 

The variance of the human P-Hem scoring could be even higher if human experts have to select a region of interest from the WSI to grade instead of getting single cut-out cells as similar research shows for the task of mitosis count~\cite{Bertram:2019vp}. We can visualise from Fig.~\ref{figHeatMap} that the score is not equally distributed over the whole image and thus the final score highly depends on the selected region of interest.
Some limitations of this work are as follows: all annotations were made by a single veterinary pathologist; data was collected at one lab, from one scanner and the data comprises of only 17 WSI. Furthermore, we have taken no action to make an external colour calibration of the participants' monitors which could positively influence the results of the participants but does not correspond to current clinical practice. In further work, we plan to analyse the effect of manual region selection by human experts and evaluate the proposed object detection pipeline to reduce its impact. Furthermore, we will introduce an interactive annotation method to increase the quality of the data set and effectively label new WSI while analysing possible bias introduced by this method. Also, it would be very interesting for us to analyse and evaluate our proposed methods and trained models on human pulmonary haemorrhage data sets. We believe transfer to human appliction may be possible using only a small data set and transfer learning.

\bibliography{sample}

\section*{Acknowledgements}

C. B. gratefully acknowledges financial support received from the Dres. Jutta \& Georg Bruns-Stiftung für innovative Veterinärmedizin.

\section*{Author contributions statement}

C. M. Created the toolchain and deep neural networks,  conceived the experiments,  analysed the results and wrote the main part of the manuscript. 
M. A. co-wrote the manuscript, created algorithmic baseline results, provided expertise through intense discussions, 
C. B. co-wrote the manuscript, Created the ground truth data set, provided expertise through intense discussions and participated in the study as expert
J. S., A. J.,F. B.,M. F.,A. B.,S. E.,S. J., J. K. Participated in the study as expert 
P. M., J. V., R. K., A. M.Provided expertise through intense discussions
J. H. Provided cytological specimens and expertise through intense discussions  All authors contributed to the preparation of the manuscript and approved of the final manuscript for publication.

\section*{Additional information}

\textbf{Competing interests.} The authors declare no competing interests.
\textbf{Data availability.} The data sets generated during and/or analysed during the current study are available from the corresponding
author on reasonable request.

\end{document}